\documentclass[final,3p,times,12pt,fixfloat]{elsarticle}
\usepackage{epsfig}
\usepackage{slashed}
\usepackage{listings}
\usepackage{booktabs,multirow,tabularx}
\usepackage{amsmath}
\usepackage{float} 
\usepackage{subfig} 
\usepackage{morefloats}
\usepackage{color}
\usepackage{multirow,bigdelim}
\usepackage{lineno}
\usepackage{tabularx}
\usepackage{subfig}

\journal{Nuclears Physics B}

\newcommand{\ben}{\begin{enumerate}}
\newcommand{\een}{\end{enumerate}}
\newcommand{\bit}{\begin{itemize}}
\newcommand{\eit}{\end{itemize}}
\newcommand{\bc}{\begin{center}}
\newcommand{\ec}{\end{center}}
\newcommand{\bq}{\begin{equation}}
\newcommand{\eq}{\end{equation}}
\newcommand{\bqa}{\begin{eqnarray}}
\newcommand{\eqa}{\end{eqnarray}}

\newcommand{\ct}[1]{{Table~(\ref{#1})}}
\newcommand{\cf}[1]{{Fig.~\ref{#1}}}

\def\p0{{\bigl.^3\hspace{-1mm}P^{[8]}_0}}

\def\to{\rightarrow}

\def\Pythia{{\sc\small Pythia}}

\def\MG5aMC{{\sc \small MadGraph5\_aMC@NLO}}
\def\MS{{\sc\small MadSpin}}

\def\MadLoop{{\sc \small MadLoop}}
\def\MadFKS{{\sc \small MadFKS}}

\usepackage{ifpdf}
\ifpdf
\usepackage[pdftex]{hyperref}
\else
\usepackage[hypertex]{hyperref}
\fi

\hypersetup{
  pdftitle={Phenomenological analysis of associated production of $Z^0+b$ in the $b \to J/\psi X$ decay channel at the LHC},%
  pdfauthor={J.P. Lansberg, H.S. Shao},%
  pdfsubject={},%
  pdfkeywords={},%
  pdfstartview={},%
  bookmarksopen=true, breaklinks=true, debug=true, %
  colorlinks=true, linkcolor=red, citecolor=blue, urlcolor=blue
}

\begin{document}

\begin{frontmatter}

\title{Phenomenological analysis of associated production of $Z^0+b$ in the $b \to J/\psi X$ decay channel at the LHC}

\author[IPNO]{Jean-Philippe Lansberg}
\author[CERN]{Hua-Sheng Shao}
\address[IPNO]{IPNO, Universit\'e Paris-Saclay, Univ. Paris-Sud, CNRS/IN2P3, F-91406, Orsay, France}
\address[CERN]{Theoretical Physics Department, CERN, CH-1211, Geneva 23, Switzerland}

\date{\today}

\begin{abstract}
The ATLAS collaboration recently reported on the first observation of associated-production
of a $Z^0$ boson with a $J/\psi$. We recently claimed that the corresponding yield
of the {\it prompt} $J/\psi$ was dominated by double parton scatterings in the ATLAS acceptance with a somewhat small
value of $\sigma_{\rm eff}$. We also found out that single parton scatterings were only dominant
at large transverse momenta. We present here the first phenomenological analysis of 
another part of the ATLAS data sample, namely of a $Z^0$ boson
plus a {\it non-prompt} $J/\psi$.  Our study is performed at next-to-leading order in $\alpha_s$ and includes 
parton-shower effects via the \MG5aMC\ framework. We find out that the data, unlike
the case of prompt $J/\psi+Z^0$, do not hint at significant DPS contributions.
Owing to the current experimental and theoretical uncertainties, there is still a room
for these but with a lower limit of $\sigma_{\rm eff}$ close to 5 mb.
We stress the importance of QCD corrections to account for the
ATLAS data.

\end{abstract}

\end{frontmatter}

\section{Introduction}

Thanks to the large luminosities of the LHC, the study of differential distributions of 
associated production of vector bosons
with open- and hidden-heavy flavour became accessible. 
These are particularly interesting because they can give access to 
 information on double parton scatterings (DPS). These,
as opposed to the conventional single parton scatterings (SPS), consist
in two simultaneous partonic scatterings during a single proton-proton collision. 

The relevance of DPS is known to increase for increasing energies at 
hadron-hadron colliders which explains why they have only started to be systematically studied
with the advent of the Tevatron and the LHC. This is also why they remain poorly understood. 
Yet, the measurement of fundamental SM parameters like the bottom-quark Yukawa 
coupling via $H^0$ and vector-boson associated production 
requires the reanalysis of these processes by taking into account DPS. 
New physics searches via same-sign $W$ boson-pair production also 
requires a good control on the DPS. 

Recent DPS studies based on quarkonium-pair production~\cite {Abazov:2014qba,Lansberg:2014swa,Abazov:2015fbl,Shao:2016wor,Aaboud:2016fzt} 
 indicate a 
smaller $\sigma_{\rm eff}$ --a parameter characterising the importance of the 
DPS ($\sigma^{\rm DPS} \propto 1/\sigma_{\rm eff}$)-- than the jet-related 
final states. We should however note that all these extractions  were carried out
under a simplified --but commonly used-- assumption whereby both scatterings	
occur independently without affecting each others' kinematics. As such,
their individual cross sections appear in a factorised way, in what we
call the "pocket formula". As of today, there do not exist proofs of 
such a factorised formula. Recent and less recent theoretical DPS studies have identified factorisation-breaking effects
 (see {\it e.g.}~\cite{Shelest:1982dg,Mekhfi:1985dv,Snigirev:2003cq,Gaunt:2009re,Blok:2010ge,Ceccopieri:2010kg,Blok:2011bu,Manohar:2012jr,Manohar:2012pe,Gaunt:2012dd,Kasemets:2012pr,Chang:2012nw,Rinaldi:2013vpa,
Diehl:2013mla,Blok:2013bpa,Diehl:2014vaa,Golec-Biernat:2014bva,Ceccopieri:2014ufa,Gaunt:2014rua,Rinaldi:2014ddl,Kasemets:2014yna,Echevarria:2015ufa,Golec-Biernat:2015aza,Diehl:2015bca})  and have thus 
shown that such a factorised "pocket formula" can only be an approximation  
and we stress that it should only be considered as such.  That being said, given the other theoretical
and experimental uncertainties involved in such extractions, such a simplification
is perfectly sound.

Quarkonia being produced via the gluon-gluon initial states, their pair 
production could help us probe the transverse correlations of the gluon-gluon 
in a proton (see {\it e.g.} \cite{Blok:2013bpa}). In fact, many quarkonium associated 
production processes have been recently measured. Let us cite 
$J/\psi$ pair production by LHCb~\cite{Aaij:2011yc}, D0~\cite{Abazov:2014qba}, 
CMS~\cite{Khachatryan:2014iia} and ATLAS~\cite{ATLAS:2016eii} ,
 $J/\psi+\Upsilon$ production by D0~\cite{Abazov:2015fbl}, 
$\Upsilon(1S)$ pair production by CMS~\cite{Khachatryan:2016ydm}, 
$J/\psi+Z^0/W^{\pm}$ production by ATLAS~\cite{Aad:2014kba,Aad:2014rua}, 
$J/\psi/\Upsilon+charm$ by LHCb~\cite{Aaij:2012dz,Aaij:2015wpa}, with their theory 
counterparts for $J/\psi+J/\psi$~\cite{Kom:2011bd,Lansberg:2013qka,Sun:2014gca,Lansberg:2014swa,Lansberg:2015lva,Baranov:2015cle,He:2015qya,Likhoded:2016zmk,Borschensky:2016nkv}, $J/\psi+\Upsilon$~\cite{Lansberg:2015lva,Shao:2016wor}, $\Upsilon+\Upsilon$~\cite{Lansberg:2015lva} and $Z^0$+ prompt $J/\psi$ \cite{Lansberg:2016rcx}. 

The observed different trend between the extracted values of 
$\sigma_{\rm eff}$ for jets and quarkonia may be the first hint  of a nontrivial flavour dependence 
of these correlations. Along these lines, the associated production of a
vector boson with heavy flavours, which we treat here, could be an unique playground to
probe corresponding quark-gluon correlations.

In this paper, we are in particular interested in the associated production 
of a $Z^0$ boson with a $b$ quark, via the observation of a non-prompt $J/\psi$, 
as measured by the ATLAS collaboration~\cite{Aad:2014kba}. This production channel is thus supposed
to probe the underlying  process $pp\rightarrow Z^0+b\bar{b}+X$\footnote{For the SPS, it proceeds via $gg\rightarrow Z^0+b\bar{b}+X$ in the 4 flavour scheme and
$g {b} \to Z^0+b$ in the 5 flavour scheme.}. At the LHC, 
such partonic reactions are usually proposed to be studied via $Z^0$ plus $b$-jet. 
We however stress that both final states are complementary since looking at
the $b$ via non-prompt $J/\psi$ allows one to access lower $P_T$ than 
with $b$-jets. For this process, we will show that going to lower $P_T$
gives the best prospects to dig out the DPS contributions, since they happen not
to be large in general. 

In addition, $Z^0+b\bar b$ production is an important observable as it can be an irreducible 
background to $Z^0+H^0$ production followed by $H \to b\bar b$, $Z^0+H^0$ being one of 
the four main $H^0$ production processes at the LHC. It could also be one 
of the crucial processes to directly probe the bottom-Higgs Yukawa coupling. 
The next-to-leading order (NLO) QCD corrections to $pp\rightarrow Z^0+b\bar{b}+X$ 
have extensively been studied in the literature~\cite{Campbell:2000bg,FebresCordero:2008ci,Cordero:2009kv,Frederix:2011qg}\footnote{The NLO electroweak corrections to the similar process $pp\rightarrow Z^0+t\bar{t}+X$ were also recently made available~\cite{Frixione:2015zaa}.}.
Yet, all the existing phenomenological studies focused on $b$-jets, hence the absence of data-theory comparisons
in~\cite{Aad:2014kba}. The present study, relying on existing and validated 
automated tools, like \MG5aMC~\cite{Alwall:2014hca} and 
\Pythia\ 8.1~\cite{Sjostrand:2007gs}, therefore fills a gap in the literature
with the first phenomenological analysis of $Z+b$ in the $b\to J/\psi X$ decay 
channel.

The structure of the article is as follows. Next section contains a short 
description of the computation set-up including the definition of the fiducial 
cuts as well as our results.  We will compare our theoretical results with the ATLAS data and extract the information of $\sigma_{\rm eff}$. Besides, a theoretical prediction will be given for the ongoing CMS measurement. Finally, we draw our conclusions in section~\ref{sec:con}.

\section{Framework and results}

\subsection{Framework}

Let us now describe how we have computed the (differential) yields to be compared
to the measurement of the ATLAS collaboration recently reported in~\cite{Aad:2014kba}.
In order to generate the (N)LO event sample for $pp\rightarrow Z^0+b\bar{b}+X$ in the 3-initial-quark-flavour scheme, we have used \MG5aMC~\cite{Alwall:2014hca}. For the record, this single framework includes \MadLoop~\cite{Hirschi:2011pa} and \MadFKS~\cite{Frederix:2009yq} to handle the virtual and real pieces respectively; the former module uses the OPP method~\cite{Ossola:2007ax,Ossola:2006us} whereas the latter uses the FKS subtraction method~\cite{Frixione:1995ms}. It also automatically uses the MC@NLO approach~\cite{Frixione:2002ik} to match NLO matrix elements to parton showers. The spin-entangled $Z^0\rightarrow e^+e^-$ decays were then performed by the \MS\ module~\cite{Artoisenet:2012st} and we have used \Pythia\ 8.1~\cite{Sjostrand:2007gs} to account for the parton showers, the hadronisation and the other decays. All this allowed us to compute the yield in the ATLAS and CMS fiducial regions (see Table~\ref{tab:phasespace}).

\begin{table*}[htpb]
\begin{center} \footnotesize
\begin{tabular}{c|c|c}
\hline\hline
\multicolumn{3}{c}{$Z$ boson selection}\\
\hline
\multicolumn{3}{c}{}\\
\multicolumn{3}{c}{$P_T$(trigger lepton)$>25$~{\rm GeV}, $P_T$(sub-leading lepton)$>15$~{\rm GeV}, $|\eta(\mathrm{lepton~from}~Z)|<2.5$}\\
\multicolumn{3}{c}{}\\
\hline\hline
\multicolumn{3}{c}{$J/\psi$ selection}\\
ATLAS fiducial~\cite{Aad:2014kba} & ATLAS inclusive~\cite{Aad:2014kba} & CMS fiducial~\cite{CMS:private}\\
\hline
$8.5<P_T^{J/\psi}<100$~{\rm GeV} & $8.5<P_T^{J/\psi}<100$~{\rm GeV} & $8.5<P_T^{J/\psi}<100$~{\rm GeV} \\
$|y_{J/\psi}|<2.1$ & $|y_{J/\psi}|<2.1$ & $|y_{J/\psi}|<2.1$\\
$P_T$(leading muon)$>4.0$~{\rm GeV} &  & $|\eta(\mathrm{muon})|<2.5$ \\
$|\eta(\mathrm{leading~muon})|<2.5$ & & \\
either \ldelim({2}{5mm}$P_T$(sub-leading muon)$>2.5$~GeV \rdelim){2}{1mm}[] & & \\
~~~~~~~~~~$1.3\leq |\eta(\textrm{sub-leading~muon})|<2.5$  & & \\
or ~~~~\ldelim({2}{5mm} $P_T$(sub-leading muon)$>3.5$~{\rm GeV} \rdelim){2}{1mm}[] &  &\\
~~~~~~~~~~$|\eta(\textrm{sub-leading~muon})|<1.3$ & & \\
\hline\hline
\end{tabular}
\caption{\label{tab:phasespace}Phase-space definition for the fiducial/inclusive production 
cross section for $J/\psi+Z$ as measured in the ATLAS detector and foreseen for the CMS detector.}
\end{center}
\end{table*}

As what regards the choice of the renormalisation scale $\mu_R$ and factorisation scale $\mu_F$, we have chosen as a central value $\mu_0=\frac{H_T}{2}$, where $H_T$ is the transverse mass sum of the final states. For the PDFs, we have used CTEQ6L1 (CTEQ6M)~\cite{Pumplin:2002vw} for the LO (NLO) computation. The integrated fragmentation fraction of $b$-hadrons to $J/\psi$ was taken to be $1.15\%$ from Ref.~\cite{Cacciari:2003uh}, which is also close to other estimations in the literature (see e.g. Refs.~\cite{Kniehl:1999vf,Bolzoni:2013tca}). The other relevant Standard Model parameters are reported in Table~\ref{tab:param}.

\begin{table*}[htpb]
\begin{center}
\begin{tabular}{cl|cl}\toprule
Parameter & Value & Parameter & Value
\\\midrule
$m_Z$ & \texttt{91.188} & $n_{lf}$ & \texttt{3}
\\
$m_{c}$ & \texttt{1.5}    & $G_{\mu}$ & \texttt{1.1987498350461625  10$^{-5}$}
\\
$m_{b}$ & \texttt{4.75}    & $\alpha_{em}^{-1}$ & \texttt{137}
\\
$m_{t}$ & \texttt{173.0}    & CKM $V_{ij}$  & $\delta_{ij}$ 
\\\bottomrule
\end{tabular}
\caption{\label{tab:param}Values of the Standard-Model parameters with the dimension-full quantity in units of GeV.}
\end{center}
\end{table*}

\subsection{Our results for the SPS contributions} 

In the ATLAS fiducial and inclusive regions (defined in Table~\ref{tab:phasespace}), we have obtained the following (N)LO SPS cross sections for {\it non-prompt} $J/\psi+Z$ production at the LHC for $\sqrt{s}=8$~TeV:	
\bqa
\sigma^{\rm LO~SPS, ATLAS~fidu.}(^{\rm np} J/\psi+Z)&=1215^{+383.5}_{-272.4}~{\rm fb}; ~ \sigma^{\rm NLO~SPS, ATLAS~fidu.}(^{\rm np}J/\psi+Z)&=1760^{+240.9}_{-220.8}~{\rm fb},\nonumber\\
\sigma^{\rm LO~SPS, ATLAS~incl.}(^{\rm np}J/\psi+Z)&=1999^{+619.7}_{-442.5}~{\rm fb};~ \sigma^{\rm NLO~SPS, ATLAS~incl.}(^{\rm np}J/\psi+Z)&=2922^{+392.9}_{-361.1}~{\rm fb},\nonumber\\
\eqa
where the theoretical uncertainty includes the renormalisation scale $\mu_R$ and factorisation scale $\mu_F$ uncertainty $\frac{\mu_0}{2}\le\mu_R,\mu_F\le 2\mu_0$, varied independently. 

\begin{figure}[hbt!]
\begin{center}
\includegraphics[width=0.45\textwidth]{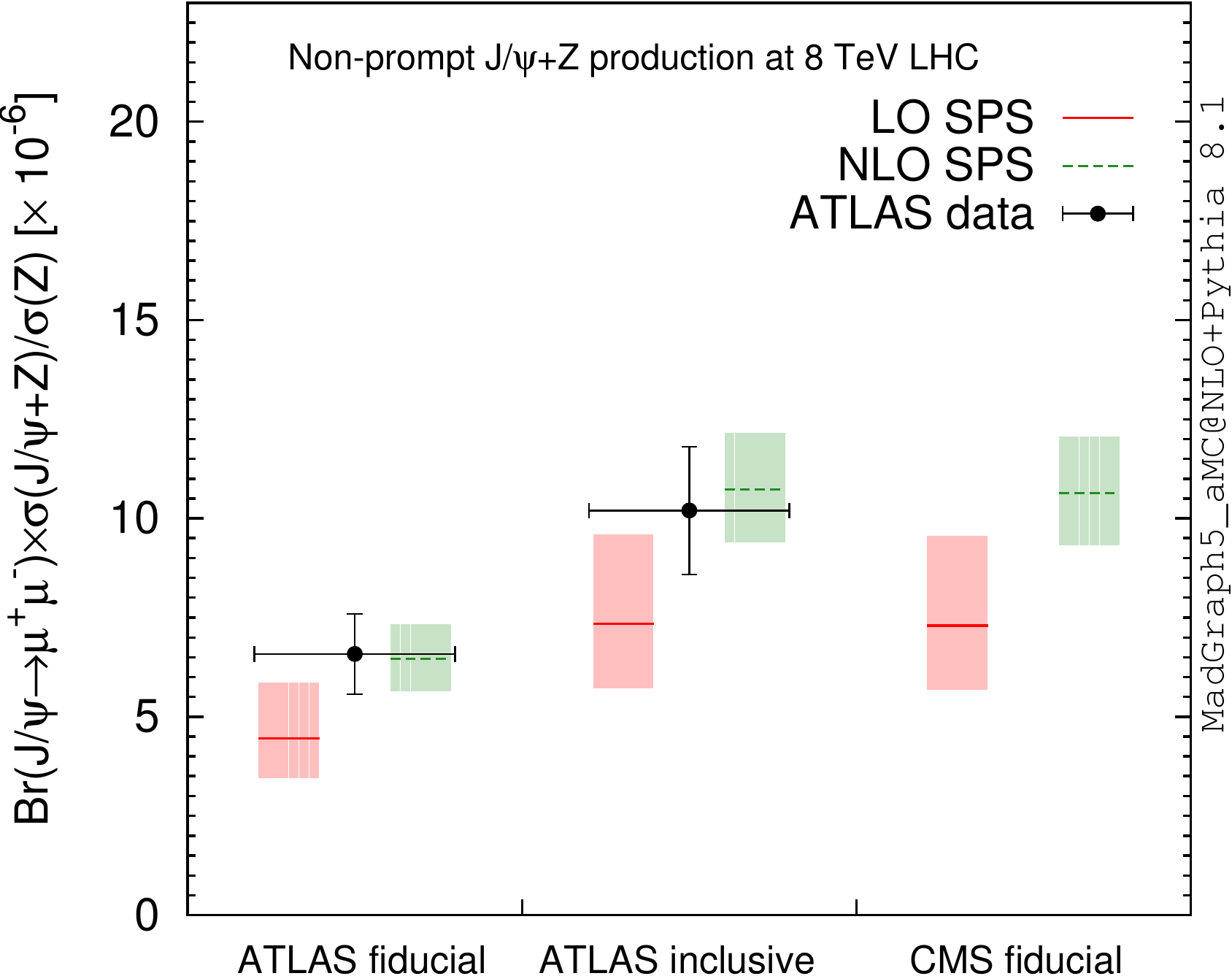}
\caption{\label{fig:totnp}Total cross section ratio $^{np}R(J/\psi+Z)$ for the non-prompt $J/\psi+Z$ production at 8 TeV LHC.}
\end{center}
\end{figure}

In absence of such predictions in the literature, no comparison was made by 
ATLAS in Ref.~\cite{Aad:2014kba}. We proceed now to the comparison to their results presented in the form 
of yield ratio in order to reduce some systematical uncertainties attached to 
the detection of the $Z$ boson, namely : 
\bqa
^{\rm np}R(J/\psi+Z)={\rm Br}(J/\psi\rightarrow \mu^+\mu^-)\times\frac{\sigma(J/\psi+Z)}{\sigma(Z)}
\eqa

We therefore  also need to use the $Z$ production cross section with 
the ATLAS cuts and have decided to simply take  
$\sigma^{\rm ATLAS}(Z)\times{\rm Br}(Z\rightarrow e^+e^-)=533.4$ fb used by ATLAS
for the comparison with the {\it prompt} $J/\psi+Z$ theory predictions.
The latter was estimated at the next-to-next-to leading order by FEWZ~\cite{Gavin:2010az}.
We have also checked this estimation with \MG5aMC\  by including the parton 
shower effects via the MC@NLO~\cite{Frixione:2002ik} approach and have obtained a 
similar value within the theory uncertainty. 

Our results and those of ATLAS are shown in Fig.\ref{fig:totnp} and 
in Table~\ref{tab:tot12}. The NLO corrections in $\alpha_s$ increase 
the $Z+$ non-prompt $J/\psi$ SPS yield by a factor of 1.46. Overall the SPS yield
ends up to be close to the ATLAS measurement and hence leaves a small room for the 
DPS yield. We also note that the relative scale uncertainty is also reduced 
from $30\%$ at LO to $13\%$ at NLO.

\begin{table}[hbt!]
\begin{center}\renewcommand{\arraystretch}{1.3}
\begin{tabularx}{\textwidth}{p{0.2\textwidth}p{0.2\textwidth}p{0.1\textwidth}p{0.12\textwidth}p{0.24\textwidth}} 
\hline\hline 
 & Experiment [$10^{-7}$] & LO SPS [$10^{-7}$] & NLO SPS [$10^{-7}$] & DPS ($\sigma_{\rm eff}=5 \div 15 $ mb) [$10^{-7}$] \\\hline
ATLAS fiducial & $65.8\pm9.2\pm4.2$ & $44.6^{+14.1}_{-10.0}$ & $64.6^{+8.8}_{-8.1}$ & -\\
ATLAS inclusive & $102\pm15\pm5\pm3$ & $73.3^{+22.7}_{-16.2}$ & $107^{+14.4}_{-13.2}$ & $8.25 \div 24.75$\\
CMS fiducial & - & $73.0^{+22.7}_{-16.2}$ & $106^{+15.3}_{-12.4}$ & - \\
\hline\hline
\end{tabularx}
\caption{\label{tab:tot12}Comparison of the cross section ratio $^{\rm np}R(J/\psi+Z)$ between the theoretical calculations and the experimental data~\cite{Aad:2014kba} at 8 TeV LHC.}
\end{center}
\end{table}

\subsection{Discussion about the DPS contributions} 

Let us now turn to the discussion of the DPS contributions. 
ATLAS~\cite{Aad:2014kba} has made an estimation of the DPS yield using the data for 
single $Z$ and non-prompt $J/\psi$ production and using the simple "pocket 
formula"\footnote{Let us recall at this stage our caveat mentioned in the introduction 
that there do not exist proofs of such a formula and that factorisation-breaking effects
have been discussed in a number of recent studies (see {\it e.g.}~\cite{Shelest:1982dg,Mekhfi:1985dv,Snigirev:2003cq,Gaunt:2009re,Blok:2010ge,Ceccopieri:2010kg,Blok:2011bu,Manohar:2012jr,Manohar:2012pe,Gaunt:2012dd,Kasemets:2012pr,Chang:2012nw,Rinaldi:2013vpa,
Diehl:2013mla,Blok:2013bpa,Diehl:2014vaa,Golec-Biernat:2014bva,Ceccopieri:2014ufa,Gaunt:2014rua,Rinaldi:2014ddl,Kasemets:2014yna,Echevarria:2015ufa,Golec-Biernat:2015aza,Diehl:2015bca}).}.
\bqa
\sigma^{\rm DPS}(J/\psi+Z)=\frac{\sigma(J/\psi)\sigma(Z)}{\sigma_{\rm eff}}.
\eqa

By assuming $\sigma_{\rm eff}=15$ mb, they quoted $^{\rm np}R(J/\psi+Z)=8.25\times 10^{-7}$ from DPS. 
If one uses a value of 5 mb, more in line with the conclusions of our study of
{\it prompt} $J/\psi+Z$~\cite{Lansberg:2016rcx}, $^{\rm np}R(J/\psi+Z)$ is thus naturally three
times as large. As evident from \ct{tab:tot12}, such a value is only marginally compatible
with the ATLAS measurements owing to the experimental and (SPS) theoretical 
uncertainties.

In fact, this also means that we can extract a lower limit on $\sigma_{\rm eff}$, 
corresponding to a maximum allowed DPS yields, now that we have at disposal 
a SPS computation. Contrary to other cases which we previously analysed~\cite{Lansberg:2014swa,Lansberg:2016rcx}, 
we cannot extract an upper limit
since the SPS yield alone is compatible with the data. 
We evaluate the $\{68\%;95\%\}$ confidence level upper limit on the SPS yield simply as follows:
\bqa
\sigma^{\rm DPS, max}=(\sigma^{\rm ATLAS~data}+\{1;2\} \times \delta \sigma^{\rm ATLAS~data})-(\sigma^{\rm SPS}-\{1;2\} \times \delta \sigma^{\rm SPS}),
\eqa	
where $\sigma$ generically denotes the central value of the $J/\psi+Z$ cross section and $\delta \sigma$ is the standard deviation of this cross section. The lower value of $\sigma_{\rm eff}$ at $68\%$ ($95\%$) confidence level is then $5.0$ mb ($2.3$ mb), which is compatible with the $\sigma_{\rm eff}$ extraction from the other quarkonium-related measurements~\cite{Abazov:2014qba,Khachatryan:2014iia,Lansberg:2014swa,Abazov:2015fbl,Shao:2016wor,Lansberg:2016rcx} and it is close to 
the range $\sigma_{\rm eff}=4.7^{+2.4}_{-1.5}$~mb we obtained for prompt $J/\psi+Z$ production~\cite{Lansberg:2016rcx}. 

\subsection{Comparison with differential distributions} 

Let us now turn to the comparison of the differential distributions between the 
theoretical results and the ATLAS data, which in fact allows us to draw similar 
conclusions. Still by lack of SPS predictions, ATLAS could only compare
its measurement of the transverse-momentum  $P_T$ spectrum of non-prompt $J/\psi$
 to their estimation of the DPS yield. As expected from the near dominance of SPS for 
this process (see above), they found out a very large discrepancy (gap between the data 
and the blue histogram). 

\begin{figure}[hbt!]
\begin{center}
\subfloat[$P_T^{J/\psi}$ differential cross section]{\includegraphics[width=0.49\textwidth]{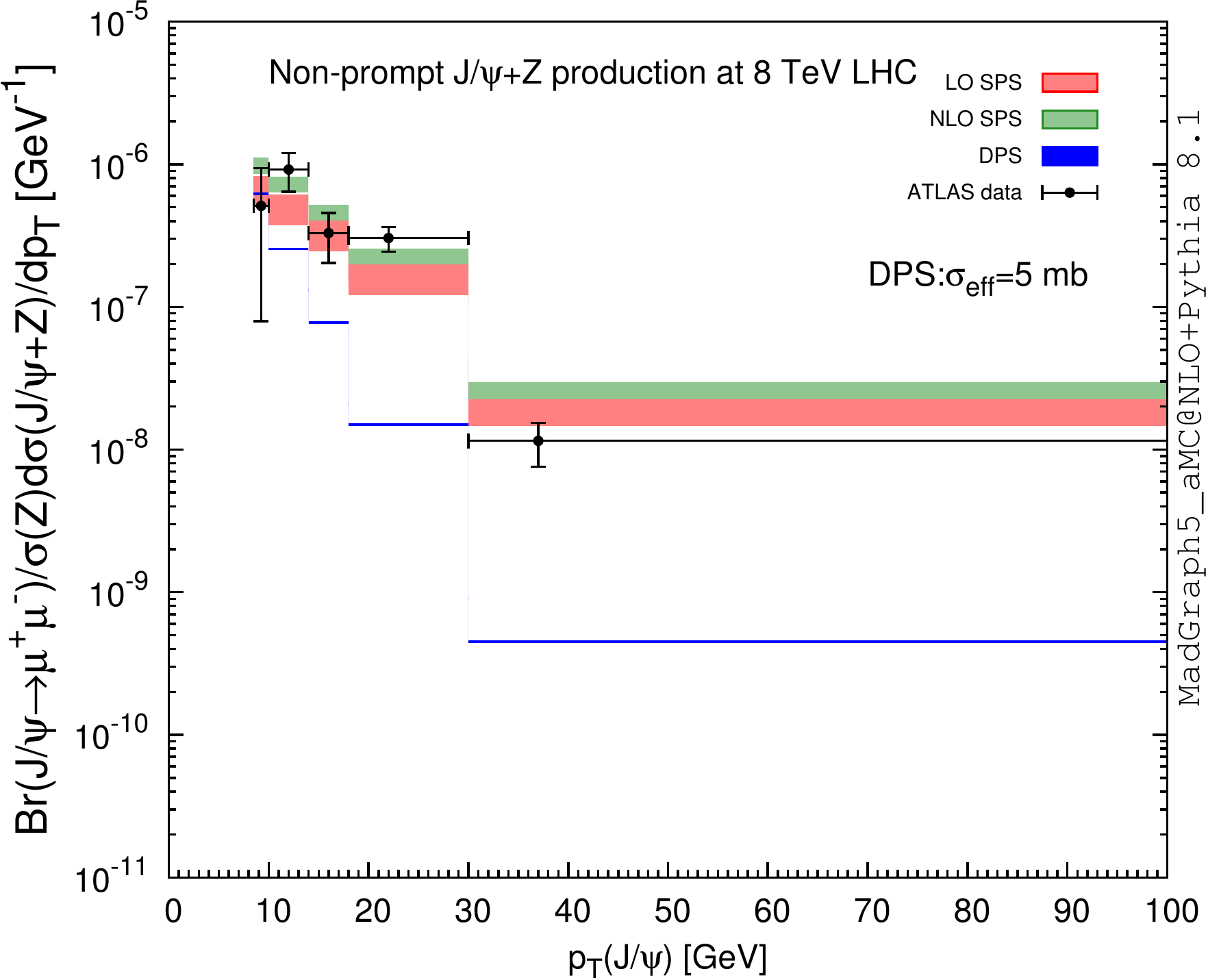}\label{fig:dpt}}
\subfloat[$\Delta \phi_{Z-J/\psi}$ differential distribution]{\includegraphics[width=0.49\textwidth]{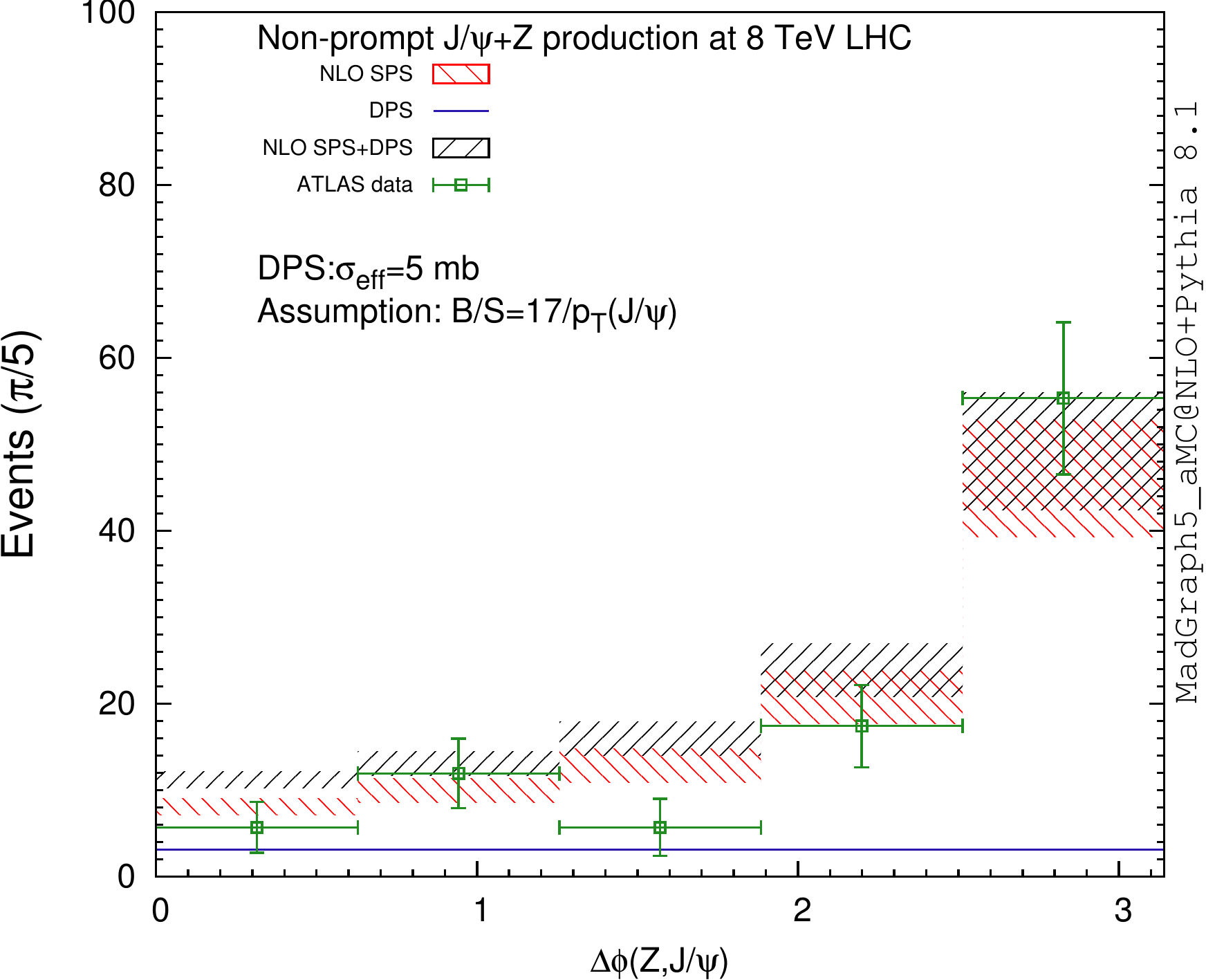}\label{fig:dphi}}
\caption{\label{fig:dpt4} Differential cross section/distributions for non-prompt $J/\psi +Z$ production: $p_T$ distribution of $J/\psi$ (a) and azimuthal angle distribution (b).}
\end{center}
\end{figure}

Adding the SPS contribution which we have computed largely fills the gap between 
theory and experiment as can be noted in \cf{fig:dpt}. Only remains a small discrepancy 
in the last $P_T^{J/\psi}$ bin  which should however be confirmed by  
forthcoming measurements as well as more accurate theoretical calculations, with {\it e.g.} an improved description of the $b$ quark fragmentation,
an account of even higher order QCD corrections, and a matching between different initial-quark flavour number schemes. Similar to the prompt $J/\psi+Z$ production, the DPS contributions  exhibit a softer $P_T^{J/\psi}$ spectrum than the SPS ones.

\begin{table}[hbt!]
\begin{center}
\begin{tabular}{{c}*2 c} 
\hline\hline
 $P_T^{J/\psi}$ [GeV] & $S$\\\hline
$(8.5,10)$ & $4.2$\\
$(10,14)$ & $32.7$\\
$(14,18)$ & $15.6$\\
$(18,30)$ & $47.1$\\
$(30,100)$ & $12.7$\\\hline
$(8.5,100)$ & $112.3$\\
\hline\hline
\end{tabular}
\caption{\label{tab:events}The estimation of the number of the signal events $S$ (before the efficiency corrections) for non-prompt $J/\psi+Z$ in each $P_T^{J/\psi}$ bin with the assumption $B/S=17/P_T^{J/\psi}$.}
\end{center}
\end{table}

Unlike the $P_T^{J/\psi}$ spectrum, ATLAS did not provide the efficiency-corrected azimuthal angle correlation between $J/\psi$ and $Z$, $\Delta \phi_{Z-J/\psi}$. Although such a distribution may significantly be  smeared by non-perturbative intrinsic initial parton $k_T$~\cite{Lansberg:2014swa,Kom:2011bd} in the low $P_T$ region, we do not think that such a smearing effect will be large here because of the ATLAS $P_T^{J/\psi}$ cut which is as large as 8 GeV. Therefore, the investigation of this correlation may indeed reveal the importance of DPS directly. However, in order to make a fair comparison, one has to unfold the efficiency since it is largely dependent on $P_T^{J/\psi}$ and hence impacts the collected number of events reported in this distribution. Along the same lines as Ref.~\cite{Lansberg:2016rcx} and thus by assuming the background events, $B$, and the signal events, $S$,  to scale like $B/S=17/P_T^{J/\psi}$, we can evaluate the number of non-prompt $J/\psi+Z$ events in each $P_T^{J/\psi}$, see \ct{tab:events}. Overall, we evaluate the total number of non-prompt signal events to be 112 vs the ATLAS found $95\pm12\pm8$, which are roughly consistent. Of course, a more precise comparison will be possible if ATLAS releases a cross section differential in $\Delta \phi_{Z-J/\psi}$. Once the number of signal events is known in each bin, one can fill the $\Delta \phi_{Z-J/\psi}$ distribution according to these numbers
using the ratio DPS/SPS in each bin as well as the expected azimuthal DPS and SPS distributions. The former is assumed to be flat
(uncorrelated production) see the blue line of \cf{fig:dphi}. The latter is obtained from our NLO computation just as the SPS $P_T^{J/\psi}$ spectrum was, see the red histogram of \cf{fig:dphi}. The resulting comparison ends up to be satisfactory and confirms that
the {\it non-prompt} $J/\psi+Z$ yield in the ATLAS acceptance is dominated by SPS contributions contrary to the {\it prompt} $J/\psi+Z$ yield~\cite{Lansberg:2016rcx}.

As aforementioned, the ratio DPS/SPS increases in the ATLAS acceptance 
for decreasing $P_T^{J/\psi}$. For the lowest $P_T^{J/\psi}$ bin, 
it even reaches 20 \% with $\sigma_{\rm eff}=15$ mb (thus 
3 times larger for 5 mb). With higher statistics, it will be possible 
to measure the yield differential in the rapidity  difference, $\Delta y$, between the $Z$ and the $J/\psi$.
For increasing, $\Delta y$, the DPS yield, {\it i.e.} from two independent scatterings, 
is favoured with respect to the SPS yield and
could eventually be dominant like for quarkonium-pair production~\cite{Kom:2011bd,Lansberg:2014swa}.

\section{Conclusions\label{sec:con}}

By providing the first phenomenological analysis of (SPS contributions to) $Z+b$ production in the $b\to J/\psi X$ decay channel, 
we have filled a gap in the literature and could make the first comparison between theory and 
the corresponding measurement by the ATLAS collaboration~\cite{Aad:2014kba}. 
Unlike the case of {\it prompt} $J/\psi+Z$ production~\cite{Lansberg:2016rcx}, we have found out that the SPS contributions to 
{\it non-prompt} $J/\psi+Z$ production happen to be dominant and very close to experimental data.
This therefore sets up an upper (lower) limit of the DPS yield ($\sigma_{\rm eff}$).

Our conclusion is based on a computation including NLO QCD corrections and parton-shower effects using \MG5aMC\ and \Pythia\ 8.1. Our comparison between the theory and the experiment also shows the importance of the QCD corrections, which not only results in a smaller scale uncertainty but also improves the agreement with data. An improved determination of $\sigma_{\rm eff}$ requires a better control on both the theoretical and the experimental uncertainties. This of course holds only if one sticks to the simple "pocket formula" where factorisation between both parton scatterings is implied as done for all the existing $\sigma_{\rm eff}$ experimental extractions.

\begin{figure}
\begin{center}
\includegraphics[width=0.75\textwidth]{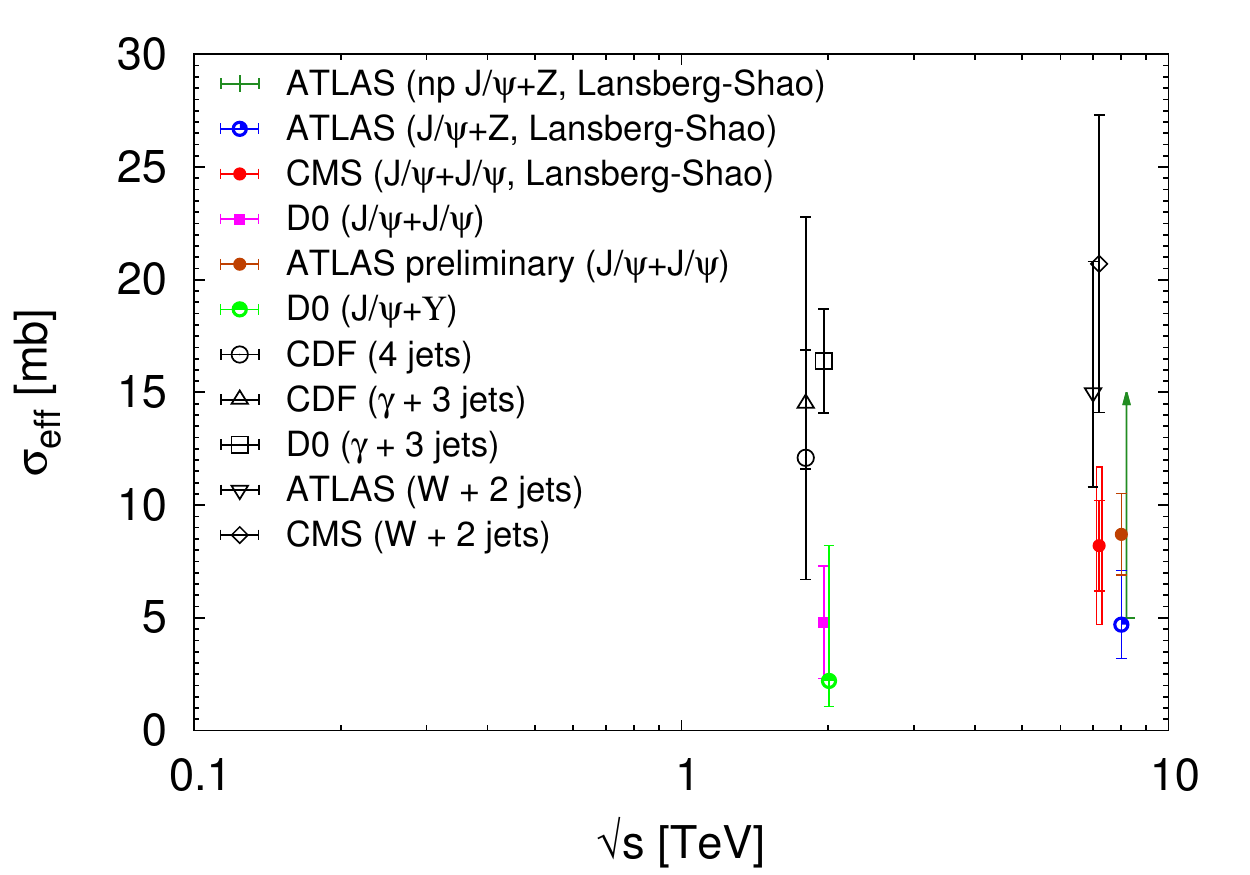}
\caption{\label{fig:sigeff} Extractions of $\sigma_{\rm eff}$ from quarkonium associated production~\cite{Lansberg:2016rcx,Lansberg:2014swa,Abazov:2014qba,ATLAS:2016eii,Abazov:2015fbl} and jet production~\cite{Aad:2013bjm,Abe:1993rv,Abe:1997xk,Abazov:2009gc,Chatrchyan:2013xxa}  processes at the Tevatron and the LHC. The symbol ``np" in the legend refers to non-prompt $J/\psi$, otherwise it refers to prompt $J/\psi$. The lower limit (green arrow) $\sigma_{\rm eff}\ge 5.0$ mb is determined from the present analysis of the ATLAS non-prompt $J/\psi+Z$ sample~\cite{Aad:2014kba}. The upper limit of $\sigma_{\rm eff}\le 8.2$ mb in D0 $J/\psi+\Upsilon$~\cite{Abazov:2015fbl} is refined in Ref.~\cite{Shao:2016wor}.}
\end{center}
\end{figure}

Based on the ATLAS measurement~\cite{Aad:2014kba} at $\sqrt{s}=8$ TeV, we thus set the lower limit of $\sigma_{\rm eff}$  to be 5.0 mb at $68\%$ confidence level and  $2.3$ mb at $95\%$ confidence level. A comparison with other extractions from both quarkonium associated production~\cite{Lansberg:2016rcx,Lansberg:2014swa,Abazov:2014qba,ATLAS:2016eii,Abazov:2015fbl} and jet production~\cite{Aad:2013bjm,Abe:1993rv,Abe:1997xk,Abazov:2009gc,Chatrchyan:2013xxa}  is displayed in \cf{fig:sigeff}. The values of $\sigma_{\rm eff}$ from quarkonium production are in general lower than those from jet production, although no strong conclusion can be drawn at the moment due to the remaining large uncertainties, some of them inherent to our incomplete knowledge of the quarkonium-production 
mechanisms~\cite{Andronic:2015wma,Brambilla:2010cs,Lansberg:2006dh}. If such an observation is confirmed in the future, it may reveal a nontrivial transverse correlations between sets of two partons from a proton or a violation of DPS factorisation, even at high energies. In general, processes as the one discussed here 
are very important because both scatterings probe different initial states --despite the expected small impact of DPS on the current data set. We emphasise that such a test is very important and, as illustrated here, is feasible at the LHC
with higher statistics which would allow one to reach parts of the phase space
where DPS contributions are expected to be larger.

\section*{Acknowledgements} 
We thank V. Kartvelishvili, D. Kikola, Y. Kubota, S. Leontsinis, F. Maltoni, D. Price, L.-P. Sun, J. Turkewitz, Z. Yang for useful discussions. 
The work of JPL is supported in part by the CNRS via FCPPL (Quarkonium4AFTER) and DEFY infiniti (THEORY LHC FRANCE). HSS is supported by the
ERC grant 291377 \textit{LHCtheory: Theoretical predictions and analyses of LHC physics:
advancing the precision frontier}.

\bibliographystyle{utphys}

\bibliography{NonPromptJpsiZ-271216}

\providecommand{\href}[2]{#2}\begingroup\raggedright\begin{thebibliography}{10}

\bibitem{Abazov:2014qba}
{\bfseries D0} Collaboration, V.~M. Abazov {\em et~al.}, ``{Observation and
  studies of double $J/\psi$ production at the Tevatron},''
  \href{http://dx.doi.org/10.1103/PhysRevD.90.111101}{{\em Phys. Rev.}
  {\bfseries D90} no.~11, (2014) 111101},
\href{http://arxiv.org/abs/1406.2380}{{\ttfamily arXiv:1406.2380 [hep-ex]}}.

\bibitem{Lansberg:2014swa}
J.-P. Lansberg and H.-S. Shao, ``{J/$\psi$ -pair production at large momenta:
  Indications for double parton scatterings and large $\alpha_s^5$
  contributions},''
  \href{http://dx.doi.org/10.1016/j.physletb.2015.10.083}{{\em Phys. Lett.}
  {\bfseries B751} (2015) 479--486},
\href{http://arxiv.org/abs/1410.8822}{{\ttfamily arXiv:1410.8822 [hep-ph]}}.

\bibitem{Abazov:2015fbl}
{\bfseries D0} Collaboration, V.~M. Abazov {\em et~al.}, ``{Evidence for
  simultaneous production of $J/\psi$ and $\Upsilon$ mesons},''
  \href{http://dx.doi.org/10.1103/PhysRevLett.116.082002}{{\em Phys. Rev.
  Lett.} {\bfseries 116} no.~8, (2016) 082002},
\href{http://arxiv.org/abs/1511.02428}{{\ttfamily arXiv:1511.02428 [hep-ex]}}.

\bibitem{Shao:2016wor}
H.-S. Shao and Y.-J. Zhang, ``{Complete study of hadroproduction of a
  $\Upsilon$ meson associated with a prompt $J/\psi$},''
  \href{http://dx.doi.org/10.1103/PhysRevLett.117.062001}{{\em Phys. Rev.
  Lett.} {\bfseries 117} no.~6, (2016) 062001},
\href{http://arxiv.org/abs/1605.03061}{{\ttfamily arXiv:1605.03061 [hep-ph]}}.

\bibitem{Aaboud:2016fzt}
{\bfseries ATLAS} Collaboration, M.~Aaboud {\em et~al.}, ``{Measurement of the
  prompt $J/\psi$ pair production cross-section in $pp$ collisions at $\sqrt{s}
  = 8$ TeV with the ATLAS detector},''
\href{http://arxiv.org/abs/1612.02950}{{\ttfamily arXiv:1612.02950 [hep-ex]}}.

\bibitem{Shelest:1982dg}
V.~P. Shelest, A.~M. Snigirev, and G.~M. Zinovev, ``{The Multiparton
  Distribution Equations in {QCD}},''
\href{http://dx.doi.org/10.1016/0370-2693(82)90049-1}{{\em Phys. Lett.}
  {\bfseries B113} (1982) 325}.

\bibitem{Mekhfi:1985dv}
M.~Mekhfi, ``{Correlations in Color and Spin in Multiparton Processes},''
\href{http://dx.doi.org/10.1103/PhysRevD.32.2380}{{\em Phys. Rev.} {\bfseries
  D32} (1985) 2380}.

\bibitem{Snigirev:2003cq}
A.~M. Snigirev, ``{Double parton distributions in the leading logarithm
  approximation of perturbative QCD},''
  \href{http://dx.doi.org/10.1103/PhysRevD.68.114012}{{\em Phys. Rev.}
  {\bfseries D68} (2003) 114012},
\href{http://arxiv.org/abs/hep-ph/0304172}{{\ttfamily arXiv:hep-ph/0304172
  [hep-ph]}}.

\bibitem{Gaunt:2009re}
J.~R. Gaunt and W.~J. Stirling, ``{Double Parton Distributions Incorporating
  Perturbative QCD Evolution and Momentum and Quark Number Sum Rules},''
  \href{http://dx.doi.org/10.1007/JHEP03(2010)005}{{\em JHEP} {\bfseries 03}
  (2010) 005},
\href{http://arxiv.org/abs/0910.4347}{{\ttfamily arXiv:0910.4347 [hep-ph]}}.

\bibitem{Blok:2010ge}
B.~Blok, {\relax Yu}.~Dokshitzer, L.~Frankfurt, and M.~Strikman, ``{The Four
  jet production at LHC and Tevatron in QCD},''
  \href{http://dx.doi.org/10.1103/PhysRevD.83.071501}{{\em Phys. Rev.}
  {\bfseries D83} (2011) 071501},
\href{http://arxiv.org/abs/1009.2714}{{\ttfamily arXiv:1009.2714 [hep-ph]}}.

\bibitem{Ceccopieri:2010kg}
F.~A. Ceccopieri, ``{An update on the evolution of double parton
  distributions},''
  \href{http://dx.doi.org/10.1016/j.physletb.2011.02.047}{{\em Phys. Lett.}
  {\bfseries B697} (2011) 482--487},
\href{http://arxiv.org/abs/1011.6586}{{\ttfamily arXiv:1011.6586 [hep-ph]}}.

\bibitem{Blok:2011bu}
B.~Blok, {\relax Yu}.~Dokshitser, L.~Frankfurt, and M.~Strikman, ``{pQCD
  physics of multiparton interactions},''
  \href{http://dx.doi.org/10.1140/epjc/s10052-012-1963-8}{{\em Eur. Phys. J.}
  {\bfseries C72} (2012) 1963},
\href{http://arxiv.org/abs/1106.5533}{{\ttfamily arXiv:1106.5533 [hep-ph]}}.

\bibitem{Manohar:2012jr}
A.~V. Manohar and W.~J. Waalewijn, ``{A QCD Analysis of Double Parton
  Scattering: Color Correlations, Interference Effects and Evolution},''
  \href{http://dx.doi.org/10.1103/PhysRevD.85.114009}{{\em Phys. Rev.}
  {\bfseries D85} (2012) 114009},
\href{http://arxiv.org/abs/1202.3794}{{\ttfamily arXiv:1202.3794 [hep-ph]}}.

\bibitem{Manohar:2012pe}
A.~V. Manohar and W.~J. Waalewijn, ``{What is Double Parton Scattering?},''
  \href{http://dx.doi.org/10.1016/j.physletb.2012.05.044}{{\em Phys. Lett.}
  {\bfseries B713} (2012) 196--201},
\href{http://arxiv.org/abs/1202.5034}{{\ttfamily arXiv:1202.5034 [hep-ph]}}.

\bibitem{Gaunt:2012dd}
J.~R. Gaunt, ``{Single Perturbative Splitting Diagrams in Double Parton
  Scattering},'' \href{http://dx.doi.org/10.1007/JHEP01(2013)042}{{\em JHEP}
  {\bfseries 01} (2013) 042},
\href{http://arxiv.org/abs/1207.0480}{{\ttfamily arXiv:1207.0480 [hep-ph]}}.

\bibitem{Kasemets:2012pr}
T.~Kasemets and M.~Diehl, ``{Angular correlations in the double Drell-Yan
  process},'' \href{http://dx.doi.org/10.1007/JHEP01(2013)121}{{\em JHEP}
  {\bfseries 01} (2013) 121},
\href{http://arxiv.org/abs/1210.5434}{{\ttfamily arXiv:1210.5434 [hep-ph]}}.

\bibitem{Chang:2012nw}
H.-M. Chang, A.~V. Manohar, and W.~J. Waalewijn, ``{Double Parton Correlations
  in the Bag Model},'' \href{http://dx.doi.org/10.1103/PhysRevD.87.034009}{{\em
  Phys. Rev.} {\bfseries D87} no.~3, (2013) 034009},
\href{http://arxiv.org/abs/1211.3132}{{\ttfamily arXiv:1211.3132 [hep-ph]}}.

\bibitem{Rinaldi:2013vpa}
M.~Rinaldi, S.~Scopetta, and V.~Vento, ``{Double parton correlations in
  constituent quark models},''
  \href{http://dx.doi.org/10.1103/PhysRevD.87.114021}{{\em Phys. Rev.}
  {\bfseries D87} (2013) 114021},
\href{http://arxiv.org/abs/1302.6462}{{\ttfamily arXiv:1302.6462 [hep-ph]}}.

\bibitem{Diehl:2013mla}
M.~Diehl and T.~Kasemets, ``{Positivity bounds on double parton
  distributions},'' \href{http://dx.doi.org/10.1007/JHEP05(2013)150}{{\em JHEP}
  {\bfseries 05} (2013) 150},
\href{http://arxiv.org/abs/1303.0842}{{\ttfamily arXiv:1303.0842 [hep-ph]}}.

\bibitem{Blok:2013bpa}
B.~Blok, {\relax Yu}.~Dokshitzer, L.~Frankfurt, and M.~Strikman,
  ``{Perturbative QCD correlations in multi-parton collisions},''
  \href{http://dx.doi.org/10.1140/epjc/s10052-014-2926-z}{{\em Eur. Phys. J.}
  {\bfseries C74} (2014) 2926},
\href{http://arxiv.org/abs/1306.3763}{{\ttfamily arXiv:1306.3763 [hep-ph]}}.

\bibitem{Diehl:2014vaa}
M.~Diehl, T.~Kasemets, and S.~Keane, ``{Correlations in double parton
  distributions: effects of evolution},''
  \href{http://dx.doi.org/10.1007/JHEP05(2014)118}{{\em JHEP} {\bfseries 05}
  (2014) 118},
\href{http://arxiv.org/abs/1401.1233}{{\ttfamily arXiv:1401.1233 [hep-ph]}}.

\bibitem{Golec-Biernat:2014bva}
K.~Golec-Biernat and E.~Lewandowska, ``{How to impose initial conditions for
  QCD evolution of double parton distributions?},''
  \href{http://dx.doi.org/10.1103/PhysRevD.90.014032}{{\em Phys. Rev.}
  {\bfseries D90} no.~1, (2014) 014032},
\href{http://arxiv.org/abs/1402.4079}{{\ttfamily arXiv:1402.4079 [hep-ph]}}.

\bibitem{Ceccopieri:2014ufa}
F.~A. Ceccopieri, ``{A second update on double parton distributions},''
  \href{http://dx.doi.org/10.1016/j.physletb.2014.05.015}{{\em Phys. Lett.}
  {\bfseries B734} (2014) 79--85},
\href{http://arxiv.org/abs/1403.2167}{{\ttfamily arXiv:1403.2167 [hep-ph]}}.

\bibitem{Gaunt:2014rua}
J.~R. Gaunt, R.~Maciula, and A.~Szczurek, ``{Conventional versus
  single-ladder-splitting contributions to double parton scattering production
  of two quarkonia, two Higgs bosons and $c \bar c c \bar c$},''
  \href{http://dx.doi.org/10.1103/PhysRevD.90.054017}{{\em Phys. Rev.}
  {\bfseries D90} no.~5, (2014) 054017},
\href{http://arxiv.org/abs/1407.5821}{{\ttfamily arXiv:1407.5821 [hep-ph]}}.

\bibitem{Rinaldi:2014ddl}
M.~Rinaldi, S.~Scopetta, M.~Traini, and V.~Vento, ``{Double parton correlations
  and constituent quark models: a Light Front approach to the valence
  sector},'' \href{http://dx.doi.org/10.1007/JHEP12(2014)028}{{\em JHEP}
  {\bfseries 12} (2014) 028},
\href{http://arxiv.org/abs/1409.1500}{{\ttfamily arXiv:1409.1500 [hep-ph]}}.

\bibitem{Kasemets:2014yna}
T.~Kasemets and P.~J. Mulders, ``{Constraining double parton correlations and
  interferences},'' \href{http://dx.doi.org/10.1103/PhysRevD.91.014015}{{\em
  Phys. Rev.} {\bfseries D91} (2015) 014015},
\href{http://arxiv.org/abs/1411.0726}{{\ttfamily arXiv:1411.0726 [hep-ph]}}.

\bibitem{Echevarria:2015ufa}
M.~G. Echevarria, T.~Kasemets, P.~J. Mulders, and C.~Pisano, ``{Polarization
  effects in double open-charm production at LHCb},''
  \href{http://dx.doi.org/10.1007/JHEP04(2015)034}{{\em JHEP} {\bfseries 04}
  (2015) 034},
\href{http://arxiv.org/abs/1501.07291}{{\ttfamily arXiv:1501.07291 [hep-ph]}}.

\bibitem{Golec-Biernat:2015aza}
K.~Golec-Biernat, E.~Lewandowska, M.~Serino, Z.~Snyder, and A.~M. Stasto,
  ``{Constraining the double gluon distribution by the single gluon
  distribution},'' \href{http://dx.doi.org/10.1016/j.physletb.2015.09.067}{{\em
  Phys. Lett.} {\bfseries B750} (2015) 559--564},
\href{http://arxiv.org/abs/1507.08583}{{\ttfamily arXiv:1507.08583 [hep-ph]}}.

\bibitem{Diehl:2015bca}
M.~Diehl, J.~R. Gaunt, D.~Ostermeier, P.~Plößl, and A.~Schäfer,
  ``{Cancellation of Glauber gluon exchange in the double Drell-Yan process},''
  \href{http://dx.doi.org/10.1007/JHEP01(2016)076}{{\em JHEP} {\bfseries 01}
  (2016) 076},
\href{http://arxiv.org/abs/1510.08696}{{\ttfamily arXiv:1510.08696 [hep-ph]}}.

\bibitem{Aaij:2011yc}
{\bfseries LHCb} Collaboration, R.~Aaij {\em et~al.}, ``{Observation of
  $J/\psi$ pair production in $pp$ collisions at $\sqrt{s}=7 TeV$},''
  \href{http://dx.doi.org/10.1016/j.physletb.2011.12.015}{{\em Phys. Lett.}
  {\bfseries B707} (2012) 52--59},
\href{http://arxiv.org/abs/1109.0963}{{\ttfamily arXiv:1109.0963 [hep-ex]}}.

\bibitem{Khachatryan:2014iia}
{\bfseries CMS} Collaboration, V.~Khachatryan {\em et~al.}, ``{Measurement of
  prompt $J/\psi$ pair production in pp collisions at $ \sqrt{s} $ = 7 Tev},''
  \href{http://dx.doi.org/10.1007/JHEP09(2014)094}{{\em JHEP} {\bfseries 09}
  (2014) 094},
\href{http://arxiv.org/abs/1406.0484}{{\ttfamily arXiv:1406.0484 [hep-ex]}}.

\bibitem{ATLAS:2016eii}
{\bfseries ATLAS} Collaboration, T.~A. collaboration,
``{Measurement of the prompt J/$\psi$ pair production cross-section in pp
  collisions at $\sqrt{s}$ = 8 TeV with the ATLAS detector},''.

\bibitem{Khachatryan:2016ydm}
{\bfseries CMS} Collaboration, V.~Khachatryan {\em et~al.}, ``{Observation of
  Upsilon(1S) pair production in proton-proton collisions at sqrt(s) = 8
  TeV},''
\href{http://arxiv.org/abs/1610.07095}{{\ttfamily arXiv:1610.07095 [hep-ex]}}.

\bibitem{Aad:2014kba}
{\bfseries ATLAS} Collaboration, G.~Aad {\em et~al.}, ``{Observation and
  measurements of the production of prompt and non-prompt $J/\psi$ mesons in
  association with a $Z$ boson in $pp$ collisions at $\sqrt{s}$ = 8 TeV with
  the ATLAS detector},''
  \href{http://dx.doi.org/10.1140/epjc/s10052-015-3406-9}{{\em Eur. Phys. J.}
  {\bfseries C75} no.~5, (2015) 229},
\href{http://arxiv.org/abs/1412.6428}{{\ttfamily arXiv:1412.6428 [hep-ex]}}.

\bibitem{Aad:2014rua}
{\bfseries ATLAS} Collaboration, G.~Aad {\em et~al.}, ``{Measurement of the
  production cross section of prompt $J/\psi$ mesons in association with a
  $W^\pm$ boson in $pp$ collisions at $\sqrt{s} =$ 7 TeV with the ATLAS
  detector},'' \href{http://dx.doi.org/10.1007/JHEP04(2014)172}{{\em JHEP}
  {\bfseries 04} (2014) 172},
\href{http://arxiv.org/abs/1401.2831}{{\ttfamily arXiv:1401.2831 [hep-ex]}}.

\bibitem{Aaij:2012dz}
{\bfseries LHCb} Collaboration, R.~Aaij {\em et~al.}, ``{Observation of double
  charm production involving open charm in pp collisions at $\sqrt{s}$ = 7
  TeV},'' \href{http://dx.doi.org/10.1007/JHEP03(2014)108,
  10.1007/JHEP06(2012)141}{{\em JHEP} {\bfseries 06} (2012) 141},
  \href{http://arxiv.org/abs/1205.0975}{{\ttfamily arXiv:1205.0975 [hep-ex]}}.
[Addendum: JHEP03,108(2014)].

\bibitem{Aaij:2015wpa}
{\bfseries LHCb} Collaboration, R.~Aaij {\em et~al.}, ``{Production of
  associated Y and open charm hadrons in pp collisions at $ \sqrt{s}=7 $ and 8
  TeV via double parton scattering},''
  \href{http://dx.doi.org/10.1007/JHEP07(2016)052}{{\em JHEP} {\bfseries 07}
  (2016) 052},
\href{http://arxiv.org/abs/1510.05949}{{\ttfamily arXiv:1510.05949 [hep-ex]}}.

\bibitem{Kom:2011bd}
C.~H. Kom, A.~Kulesza, and W.~J. Stirling, ``{Pair Production of J/psi as a
  Probe of Double Parton Scattering at LHCb},''
  \href{http://dx.doi.org/10.1103/PhysRevLett.107.082002}{{\em Phys. Rev.
  Lett.} {\bfseries 107} (2011) 082002},
\href{http://arxiv.org/abs/1105.4186}{{\ttfamily arXiv:1105.4186 [hep-ph]}}.

\bibitem{Lansberg:2013qka}
J.-P. Lansberg and H.-S. Shao, ``{Production of $J/\psi + \eta_{c}$ versus
  $J/\psi + J/\psi$ at the LHC: Importance of Real $\alpha^{5}_{s}$
  Corrections},'' \href{http://dx.doi.org/10.1103/PhysRevLett.111.122001}{{\em
  Phys. Rev. Lett.} {\bfseries 111} (2013) 122001},
\href{http://arxiv.org/abs/1308.0474}{{\ttfamily arXiv:1308.0474 [hep-ph]}}.

\bibitem{Sun:2014gca}
L.-P. Sun, H.~Han, and K.-T. Chao, ``{Impact of $J/\psi$ pair production at the
  LHC and predictions in nonrelativistic QCD},''
  \href{http://dx.doi.org/10.1103/PhysRevD.94.074033}{{\em Phys. Rev.}
  {\bfseries D94} no.~7, (2016) 074033},
\href{http://arxiv.org/abs/1404.4042}{{\ttfamily arXiv:1404.4042 [hep-ph]}}.

\bibitem{Lansberg:2015lva}
J.-P. Lansberg and H.-S. Shao, ``{Double-quarkonium production at a
  fixed-target experiment at the LHC (AFTER@LHC)},''
  \href{http://dx.doi.org/10.1016/j.nuclphysb.2015.09.005}{{\em Nucl. Phys.}
  {\bfseries B900} (2015) 273--294},
\href{http://arxiv.org/abs/1504.06531}{{\ttfamily arXiv:1504.06531 [hep-ph]}}.

\bibitem{Baranov:2015cle}
S.~P. Baranov and A.~H. Rezaeian, ``{Prompt double $J/\psi$ production in
  proton-proton collisions at the LHC},''
  \href{http://dx.doi.org/10.1103/PhysRevD.93.114011}{{\em Phys. Rev.}
  {\bfseries D93} no.~11, (2016) 114011},
\href{http://arxiv.org/abs/1511.04089}{{\ttfamily arXiv:1511.04089 [hep-ph]}}.

\bibitem{He:2015qya}
Z.-G. He and B.~A. Kniehl, ``{Complete Nonrelativistic-QCD Prediction for
  Prompt Double J/? Hadroproduction},''
  \href{http://dx.doi.org/10.1103/PhysRevLett.115.022002}{{\em Phys. Rev.
  Lett.} {\bfseries 115} no.~2, (2015) 022002},
\href{http://arxiv.org/abs/1609.02786}{{\ttfamily arXiv:1609.02786 [hep-ph]}}.

\bibitem{Likhoded:2016zmk}
A.~K. Likhoded, A.~V. Luchinsky, and S.~V. Poslavsky, ``{Production of $J/\psi
  + \chi_c$ and $J/\psi + J/\psi$ with real gluon emission at LHC},''
  \href{http://dx.doi.org/10.1103/PhysRevD.94.054017}{{\em Phys. Rev.}
  {\bfseries D94} no.~5, (2016) 054017},
\href{http://arxiv.org/abs/1606.06767}{{\ttfamily arXiv:1606.06767 [hep-ph]}}.

\bibitem{Borschensky:2016nkv}
C.~Borschensky and A.~Kulesza, ``{Double parton scattering in pair-production
  of $J/\psi$ mesons at the LHC revisited},''
\href{http://arxiv.org/abs/1610.00666}{{\ttfamily arXiv:1610.00666 [hep-ph]}}.

\bibitem{Lansberg:2016rcx}
J.-P. Lansberg and H.-S. Shao, ``{Associated production of a quarkonium and a Z
  boson at one loop in a quark-hadron-duality approach},''
  \href{http://dx.doi.org/10.1007/JHEP10(2016)153}{{\em JHEP} {\bfseries 10}
  (2016) 153},
\href{http://arxiv.org/abs/1608.03198}{{\ttfamily arXiv:1608.03198 [hep-ph]}}.

\bibitem{Campbell:2000bg}
J.~M. Campbell and R.~K. Ellis, ``{Radiative corrections to Z b anti-b
  production},'' \href{http://dx.doi.org/10.1103/PhysRevD.62.114012}{{\em Phys.
  Rev.} {\bfseries D62} (2000) 114012},
\href{http://arxiv.org/abs/hep-ph/0006304}{{\ttfamily arXiv:hep-ph/0006304
  [hep-ph]}}.

\bibitem{FebresCordero:2008ci}
F.~Febres~Cordero, L.~Reina, and D.~Wackeroth, ``{NLO QCD corrections to $Z b
  \bar{b}$ production with massive bottom quarks at the Fermilab Tevatron},''
  \href{http://dx.doi.org/10.1103/PhysRevD.78.074014}{{\em Phys. Rev.}
  {\bfseries D78} (2008) 074014},
\href{http://arxiv.org/abs/0806.0808}{{\ttfamily arXiv:0806.0808 [hep-ph]}}.

\bibitem{Cordero:2009kv}
F.~Febres~Cordero, L.~Reina, and D.~Wackeroth, ``{W- and Z-boson production
  with a massive bottom-quark pair at the Large Hadron Collider},''
  \href{http://dx.doi.org/10.1103/PhysRevD.80.034015}{{\em Phys. Rev.}
  {\bfseries D80} (2009) 034015},
\href{http://arxiv.org/abs/0906.1923}{{\ttfamily arXiv:0906.1923 [hep-ph]}}.

\bibitem{Frederix:2011qg}
R.~Frederix, S.~Frixione, V.~Hirschi, F.~Maltoni, R.~Pittau, and P.~Torrielli,
  ``{W and $Z/\gamma*$ boson production in association with a bottom-antibottom
  pair},'' \href{http://dx.doi.org/10.1007/JHEP09(2011)061}{{\em JHEP}
  {\bfseries 09} (2011) 061},
\href{http://arxiv.org/abs/1106.6019}{{\ttfamily arXiv:1106.6019 [hep-ph]}}.

\bibitem{Frixione:2015zaa}
S.~Frixione, V.~Hirschi, D.~Pagani, H.~S. Shao, and M.~Zaro, ``{Electroweak and
  QCD corrections to top-pair hadroproduction in association with heavy
  bosons},'' \href{http://dx.doi.org/10.1007/JHEP06(2015)184}{{\em JHEP}
  {\bfseries 06} (2015) 184},
\href{http://arxiv.org/abs/1504.03446}{{\ttfamily arXiv:1504.03446 [hep-ph]}}.

\bibitem{Alwall:2014hca}
J.~Alwall, R.~Frederix, S.~Frixione, V.~Hirschi, F.~Maltoni, O.~Mattelaer,
  H.~S. Shao, T.~Stelzer, P.~Torrielli, and M.~Zaro, ``{The automated
  computation of tree-level and next-to-leading order differential cross
  sections, and their matching to parton shower simulations},''
  \href{http://dx.doi.org/10.1007/JHEP07(2014)079}{{\em JHEP} {\bfseries 07}
  (2014) 079},
\href{http://arxiv.org/abs/1405.0301}{{\ttfamily arXiv:1405.0301 [hep-ph]}}.

\bibitem{Sjostrand:2007gs}
T.~Sjostrand, S.~Mrenna, and P.~Z. Skands, ``{A Brief Introduction to PYTHIA
  8.1},'' \href{http://dx.doi.org/10.1016/j.cpc.2008.01.036}{{\em Comput. Phys.
  Commun.} {\bfseries 178} (2008) 852--867},
\href{http://arxiv.org/abs/0710.3820}{{\ttfamily arXiv:0710.3820 [hep-ph]}}.

\bibitem{Hirschi:2011pa}
V.~Hirschi, R.~Frederix, S.~Frixione, M.~V. Garzelli, F.~Maltoni, and
  R.~Pittau, ``{Automation of one-loop QCD corrections},''
  \href{http://dx.doi.org/10.1007/JHEP05(2011)044}{{\em JHEP} {\bfseries 05}
  (2011) 044},
\href{http://arxiv.org/abs/1103.0621}{{\ttfamily arXiv:1103.0621 [hep-ph]}}.

\bibitem{Frederix:2009yq}
R.~Frederix, S.~Frixione, F.~Maltoni, and T.~Stelzer, ``{Automation of
  next-to-leading order computations in QCD: The FKS subtraction},''
  \href{http://dx.doi.org/10.1088/1126-6708/2009/10/003}{{\em JHEP} {\bfseries
  10} (2009) 003},
\href{http://arxiv.org/abs/0908.4272}{{\ttfamily arXiv:0908.4272 [hep-ph]}}.

\bibitem{Ossola:2007ax}
G.~Ossola, C.~G. Papadopoulos, and R.~Pittau, ``{CutTools: A Program
  implementing the OPP reduction method to compute one-loop amplitudes},''
  \href{http://dx.doi.org/10.1088/1126-6708/2008/03/042}{{\em JHEP} {\bfseries
  03} (2008) 042},
\href{http://arxiv.org/abs/0711.3596}{{\ttfamily arXiv:0711.3596 [hep-ph]}}.

\bibitem{Ossola:2006us}
G.~Ossola, C.~G. Papadopoulos, and R.~Pittau, ``{Reducing full one-loop
  amplitudes to scalar integrals at the integrand level},''
  \href{http://dx.doi.org/10.1016/j.nuclphysb.2006.11.012}{{\em Nucl. Phys.}
  {\bfseries B763} (2007) 147--169},
\href{http://arxiv.org/abs/hep-ph/0609007}{{\ttfamily arXiv:hep-ph/0609007
  [hep-ph]}}.

\bibitem{Frixione:1995ms}
S.~Frixione, Z.~Kunszt, and A.~Signer, ``{Three jet cross-sections to
  next-to-leading order},''
  \href{http://dx.doi.org/10.1016/0550-3213(96)00110-1}{{\em Nucl. Phys.}
  {\bfseries B467} (1996) 399--442},
\href{http://arxiv.org/abs/hep-ph/9512328}{{\ttfamily arXiv:hep-ph/9512328
  [hep-ph]}}.

\bibitem{Frixione:2002ik}
S.~Frixione and B.~R. Webber, ``{Matching NLO QCD computations and parton
  shower simulations},''
  \href{http://dx.doi.org/10.1088/1126-6708/2002/06/029}{{\em JHEP} {\bfseries
  06} (2002) 029},
\href{http://arxiv.org/abs/hep-ph/0204244}{{\ttfamily arXiv:hep-ph/0204244
  [hep-ph]}}.

\bibitem{Artoisenet:2012st}
P.~Artoisenet, R.~Frederix, O.~Mattelaer, and R.~Rietkerk, ``{Automatic
  spin-entangled decays of heavy resonances in Monte Carlo simulations},''
  \href{http://dx.doi.org/10.1007/JHEP03(2013)015}{{\em JHEP} {\bfseries 03}
  (2013) 015},
\href{http://arxiv.org/abs/1212.3460}{{\ttfamily arXiv:1212.3460 [hep-ph]}}.

\bibitem{CMS:private}
Y.~Kubota,
``{private communication},''.

\bibitem{Pumplin:2002vw}
J.~Pumplin, D.~R. Stump, J.~Huston, H.~L. Lai, P.~M. Nadolsky, and W.~K. Tung,
  ``{New generation of parton distributions with uncertainties from global QCD
  analysis},'' \href{http://dx.doi.org/10.1088/1126-6708/2002/07/012}{{\em
  JHEP} {\bfseries 07} (2002) 012},
\href{http://arxiv.org/abs/hep-ph/0201195}{{\ttfamily arXiv:hep-ph/0201195
  [hep-ph]}}.

\bibitem{Cacciari:2003uh}
M.~Cacciari, S.~Frixione, M.~L. Mangano, P.~Nason, and G.~Ridolfi, ``{QCD
  analysis of first $b$ cross-section data at 1.96-TeV},''
  \href{http://dx.doi.org/10.1088/1126-6708/2004/07/033}{{\em JHEP} {\bfseries
  07} (2004) 033},
\href{http://arxiv.org/abs/hep-ph/0312132}{{\ttfamily arXiv:hep-ph/0312132
  [hep-ph]}}.

\bibitem{Kniehl:1999vf}
B.~A. Kniehl and G.~Kramer, ``{Inclusive $J/\psi$ and $\psi_{2S}$ production
  from $B$ decay in $p \bar{p}$ collisions},''
  \href{http://dx.doi.org/10.1103/PhysRevD.60.014006}{{\em Phys. Rev.}
  {\bfseries D60} (1999) 014006},
\href{http://arxiv.org/abs/hep-ph/9901348}{{\ttfamily arXiv:hep-ph/9901348
  [hep-ph]}}.

\bibitem{Bolzoni:2013tca}
P.~Bolzoni, B.~A. Kniehl, and G.~Kramer, ``{Inclusive J/psi and psi(2S)
  production from b-hadron decay in p anti-p and pp collisions},''
  \href{http://dx.doi.org/10.1103/PhysRevD.88.074035}{{\em Phys. Rev.}
  {\bfseries D88} no.~7, (2013) 074035},
\href{http://arxiv.org/abs/1309.3389}{{\ttfamily arXiv:1309.3389 [hep-ph]}}.

\bibitem{Gavin:2010az}
R.~Gavin, Y.~Li, F.~Petriello, and S.~Quackenbush, ``{FEWZ 2.0: A code for
  hadronic Z production at next-to-next-to-leading order},''
  \href{http://dx.doi.org/10.1016/j.cpc.2011.06.008}{{\em Comput. Phys.
  Commun.} {\bfseries 182} (2011) 2388--2403},
\href{http://arxiv.org/abs/1011.3540}{{\ttfamily arXiv:1011.3540 [hep-ph]}}.

\bibitem{Aad:2013bjm}
{\bfseries ATLAS} Collaboration, G.~Aad {\em et~al.}, ``{Measurement of hard
  double-parton interactions in $W(\to l\nu)$+ 2 jet events at $\sqrt{s}$=7 TeV
  with the ATLAS detector},''
  \href{http://dx.doi.org/10.1088/1367-2630/15/3/033038}{{\em New J. Phys.}
  {\bfseries 15} (2013) 033038},
\href{http://arxiv.org/abs/1301.6872}{{\ttfamily arXiv:1301.6872 [hep-ex]}}.

\bibitem{Abe:1993rv}
{\bfseries CDF} Collaboration, F.~Abe {\em et~al.}, ``{Study of four jet events
  and evidence for double parton interactions in $p\bar{p}$ collisions at
  $\sqrt{s} = 1.8$ TeV},''
\href{http://dx.doi.org/10.1103/PhysRevD.47.4857}{{\em Phys. Rev.} {\bfseries
  D47} (1993) 4857--4871}.

\bibitem{Abe:1997xk}
{\bfseries CDF} Collaboration, F.~Abe {\em et~al.}, ``{Double parton scattering
  in $\bar{p}p$ collisions at $\sqrt{s} = 1.8 $TeV},''
\href{http://dx.doi.org/10.1103/PhysRevD.56.3811}{{\em Phys. Rev.} {\bfseries
  D56} (1997) 3811--3832}.

\bibitem{Abazov:2009gc}
{\bfseries D0} Collaboration, V.~M. Abazov {\em et~al.}, ``{Double parton
  interactions in $\gamma$+3 jet events in $p p^-$ bar collisions
  $\sqrt{s}=1.96$ TeV.},''
  \href{http://dx.doi.org/10.1103/PhysRevD.81.052012}{{\em Phys. Rev.}
  {\bfseries D81} (2010) 052012},
\href{http://arxiv.org/abs/0912.5104}{{\ttfamily arXiv:0912.5104 [hep-ex]}}.

\bibitem{Chatrchyan:2013xxa}
{\bfseries CMS} Collaboration, S.~Chatrchyan {\em et~al.}, ``{Study of double
  parton scattering using W + 2-jet events in proton-proton collisions at
  $\sqrt{s}$ = 7 TeV},'' \href{http://dx.doi.org/10.1007/JHEP03(2014)032}{{\em
  JHEP} {\bfseries 03} (2014) 032},
\href{http://arxiv.org/abs/1312.5729}{{\ttfamily arXiv:1312.5729 [hep-ex]}}.

\bibitem{Andronic:2015wma}
A.~Andronic {\em et~al.}, ``{Heavy-flavour and quarkonium production in the LHC
  era: from proton-proton to heavy-ion collisions},''
  \href{http://dx.doi.org/10.1140/epjc/s10052-015-3819-5}{{\em Eur. Phys. J.}
  {\bfseries C76} no.~3, (2016) 107},
\href{http://arxiv.org/abs/1506.03981}{{\ttfamily arXiv:1506.03981 [nucl-ex]}}.

\bibitem{Brambilla:2010cs}
N.~Brambilla {\em et~al.}, ``{Heavy quarkonium: progress, puzzles, and
  opportunities},''
  \href{http://dx.doi.org/10.1140/epjc/s10052-010-1534-9}{{\em Eur. Phys. J.}
  {\bfseries C71} (2011) 1534},
\href{http://arxiv.org/abs/1010.5827}{{\ttfamily arXiv:1010.5827 [hep-ph]}}.

\bibitem{Lansberg:2006dh}
J.~P. Lansberg, ``{$J/\psi$, $\psi$ ' and $\Upsilon$ production at hadron
  colliders: A Review},''
  \href{http://dx.doi.org/10.1142/S0217751X06033180}{{\em Int. J. Mod. Phys.}
  {\bfseries A21} (2006) 3857--3916},
\href{http://arxiv.org/abs/hep-ph/0602091}{{\ttfamily arXiv:hep-ph/0602091
  [hep-ph]}}.

\end{thebibliography}\endgroup

\end{document}